\documentclass[prl, superscriptaddress, floatfix,twocolumn,showpacs,amsfonts,amsmath,amssymb,final,superscriptaddress]{revtex4}

\usepackage{graphicx}
\usepackage{rotating}
\usepackage{amsmath}
\usepackage{amsfonts}
\usepackage{amssymb}
\usepackage[countmax]{subfloat}
\usepackage{dcolumn} 
\usepackage{bm} 
\usepackage{color}

\usepackage{float}
\usepackage{latexsym}
\usepackage{stmaryrd}

\begin{document}

\title{Minimal Model of Spin-Transfer Torque and Spin Pumping caused by Spin Hall Effect}







\author{Wei Chen}  

\affiliation{Theoretische Physik, ETH-Z\"urich, CH-8093 Z\"urich, Switzerland}

\affiliation{Max-Planck-Institut f$\ddot{u}$r Festk$\ddot{o}$rperforschung, Heisenbergstrasse 1, D-70569 Stuttgart, Germany}

\author{Manfred Sigrist}

\affiliation{Theoretische Physik, ETH-Z\"urich, CH-8093 Z\"urich, Switzerland}

\author{Jairo Sinova}

\affiliation{Department of Physics, Texas A\&M University, College Station, Texas 77843-4242, USA}

\affiliation{Institut f\"{u}r Physik, Johannes Gutenberg-Universit\"{a}t Mainz, 55128 Mainz, Germany}

\affiliation{Institute of Physics ASCR, Cukrovarnick\'{a} 10, 162 53 Praha 6, Czech Republic}

\author{Dirk Manske}

\affiliation{Max-Planck-Institut f$\ddot{u}$r Festk$\ddot{o}$rperforschung, Heisenbergstrasse 1, D-70569 Stuttgart, Germany}

\date{\rm\today}

\begin{abstract}

{In the normal metal/ferromagnetic insulator bilayer (such as Pt/Y$_{3}$Fe$_{5}$O$_{12}$) and the normal metal/ferromagnetic metal/oxide trilayer (such as Pt/Co/AlO$_{x}$) where spin injection and ejection are achieved by the spin Hall effect in the normal metal, we propose a minimal model based on quantum tunneling of spins to explain the spin-transfer torque and spin pumping caused by the spin Hall effect. The ratio of their damping-like to field-like component depends on the tunneling wave function that is strongly influenced by generic material properties such as interface $s-d$ coupling, insulating gap, and layer thickness, yet the spin relaxation plays a minor role.  The quantified result renders our minimal model an inexpensive tool for searching for appropriate materials. }

\end{abstract}





\maketitle

{\it Introduction.-} Two reciprocal mechanisms, namely the spin-transfer torque (STT)\cite{Berger96,Slonczewski96} and spin pumping\cite{Tserkovnyak02,Zhang04}, play the major roles in spintronic devices. Besides in the originally proposed metallic heterostructures, they also manifest in heterostructures involving insulators, such as the normal metal/ferromagnetic insulator (NM/FMI) bilayer realized by Pt/Y$_{3}$Fe$_{5}$O$_{12}$ (Pt/YIG)\cite{Kajiwara10}, where STT can be used to excite magnons\cite{Xiao12,Zhou13}. Unlike in metallic heterostructures, the usual way of spin injection by using a spin-polarized charge current in the current-perpendicular-to-plane (CPP) geometry is difficult in NM/FMI because the insulating FMI impedes the charge current. Instead, the spin injection in this system is done by using the spin Hall effect (SHE) in the NM, which in the current-in-plane (CIP) geometry injects a pure spin current into the FMI to cause STT. In the reciprocal process, SHE converts the pure spin current generated from spin pumping into electric signals. Despite pointing at promising applications in magnetic memory devices, the microscopic mechanism concerning how the pure spin current tunnels into an insulator to cause STT and spin pumping remains unclear, especially if spin relaxation plays a crucial role as in metallic heterostructures\cite{Berger96}.

Another system of similar kind yet involves more complexity is the normal metal/ferromagnetic metal/oxide (NM/FMM/oxide) trilayer, such as Pt/Co/AlO$_{x}$ and similar ones\cite{Miron10,Miron11,Liu12,Liu12_2,Garello13}. It is generally suspected that the observed spin torque contains both SHE-STT\cite{Liu12_2} and the spin-orbit torque (SOT)\cite{Manchon08,Haney10,Pesin12,Haney13} that stems from the charge current flowing through the parity-breaking FMM. Concerning SOT alone, a field-like component is expected from the current induced spin density\cite{Gambardella11}, yet scattering-related mechanisms\cite{Kim12,Pesin12,Haney13,Wang12}, as well as intrinsic Berry curvature\cite{Kurebayashi14} have been shown to cause a damping-like torque of similar strength. On the other hand, the mechanism of SHE-STT is less explored, thus the relative weighting between SOT and SHE-STT remains unclear. Moreover, it is technically important to single out the contribution from SHE-STT alone and investigate if this part can be enhanced by any means.

In this Letter, we present a minimal model for the STT and spin pumping in NM/FMI and NM/FMM/oxide based on quantum tunneling of spins. We point out that the damping-like and field-like component (defined in Eq.~(\ref{effective_EOM})) depend on the wave function tunneled into the ferromagnet that is strongly influenced by generic material properties such as interface $s-d$ coupling, insulating gap, and thickness of the ferromagnet, but not crucially on spin relaxation. Using the two methods that originally predict the STT in metallic heterostructures, namely the Landau-Lifshitz (LL) dynamics\cite{Berger96} and angular momentum conservation\cite{Slonczewski96}, the spin mixing conductance\cite{Maekawa12} for STT is calculated. At present, it is of particular interest to find ways to enhance the spin mixing conductance, which may realize magnetic memory devices with more efficient magnetization switching and lower power consumption, and our minimal model serves as an inexpensive tool to guide the search for suitable materials. The mechanism of spin pumping is further clarified based on the adiabatic assumption, yielding Onsager relation explicitly satisfied within this minimal model. The consistency with various experiments will be explained thoroughly.

\begin{figure}[ht]
\begin{center}
\includegraphics[clip=true,width=0.95\columnwidth]{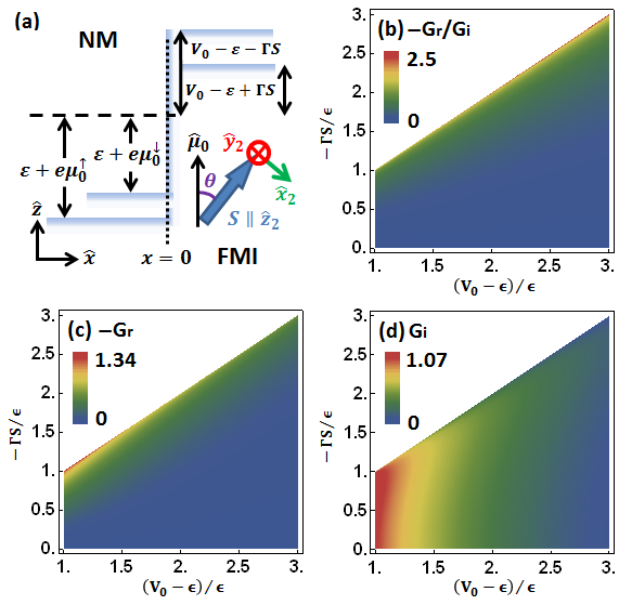}
\caption{(color online) (a) Definition of the coordinates at $\varphi=0$ and energy scales near the NM/FMI interface. Blue line indicates the potential profile seen by the conduction electrons. (b) The ratio of (c) damping-like to (d) field-like component of the spin mixing conductance at $\theta=0.3\pi$ plotted as a function of the insulating gap $\left(V_{0}-\epsilon\right)/\epsilon$ and interface $s-d$ coupling $-\Gamma S/\epsilon$ relative to the Fermi energy. The color scale is in units of $e^{2}/\hbar a^{2}$ which is about $10^{-15}\sim 10^{-14}\Omega^{-1}$m$^{-2}$ depending on the lattice constant $a$. The white regions in (b)$\sim$(d) are where one potential barrier is lower than the Fermi energy $V_{0}-\epsilon+\Gamma S<\epsilon$, hence irrelevant to the problem. If the effective mass in NM and in FMI are different, the axis labels of the plots are replaced by $-\Gamma S/\epsilon\rightarrow -m_{F}\Gamma S/m_{N}\epsilon$ and $\left(V_{0}-\epsilon\right)/\epsilon\rightarrow m_{F}\left(V_{0}-\epsilon\right)/m_{N}\epsilon$. } 
\label{fig:fitting_Gr_gi_alphap}
\end{center}
\end{figure}

{\it NM/FMI bilayer.-} We first address the quantum tunneling of spins in the NM/FMI bilayer. Consider the bilayer shown in Fig.~\ref{fig:fitting_Gr_gi_alphap} (a) that contains two regions: (1) A NM at $-\infty<x<0$ described semiclasically by $H_{N}=p^{2}/2m-\mu_{x}^{\sigma}$, where $\mu_{x}^{\sigma}=\pm|{\boldsymbol\mu}_{x}|/2$ is the spin voltage of spin $\sigma=\left\{\uparrow,\downarrow\right\}$ at position $x$ caused by SHE\cite{Zhang00,Takahashi06,Kato04}. For an up spin incident from the left, the wave function at $x<0$ is
\begin{eqnarray}
\psi_{N}=\left(Ae^{ik_{0\uparrow}x}
+Be^{-ik_{0\uparrow}x}\right)\left(
\begin{array}{l}
1 \\
0
\end{array}
\right)
+Ce^{-ik_{0\downarrow}x}\left(
\begin{array}{l}
0 \\
1
\end{array}
\right),
\label{NM_wave_function}
\end{eqnarray}
where $k_{0\sigma}=\sqrt{2m\left(\epsilon+\mu_{0}^{\sigma}\right)}/\hbar$, and $\epsilon$ is the Fermi energy. We consider a charge current $j_{y}^{c}{\hat{\bf y}}$, so electrons moving in ${\hat{\bf x}}$ direction has ${\boldsymbol\sigma}\parallel{\hat{\bf z}}$ because of the SHE relation $k_{0\sigma}{\hat{\bf x}}\parallel{\boldsymbol\sigma}\times{\hat{\bf y}}$, and gives a positive spin voltage ${\boldsymbol\mu}_{0}\parallel{\hat{\bf z}}$ at the interface such that $k_{0\uparrow}>k_{0\downarrow}$. (2) A FMI at $x\geq 0$ described by
$H_{FI}=p^{2}/2m+V_{0}+\Gamma{\bf S}\cdot{\boldsymbol \sigma}$, 
where $V_{0}>\epsilon$ is the potential step. The ${\bf S}\cdot{\boldsymbol \sigma}$ term describes the $s-d$ hybridization of the conduction electron spin and the magnetization ${\bf S}=S(\sin\theta\cos\varphi,\sin\theta\sin\varphi,\cos\theta)$. The evanescent wave function in the FMI is 
\begin{eqnarray}
\psi_{FI}&=&De^{-q_{+}x}
\left(
\begin{array}{l}
e^{-i\varphi/2}\cos\frac{\theta}{2} \\
e^{i\varphi/2}\sin\frac{\theta}{2} 
\end{array}
\right)
\nonumber \\
&&+Ee^{-q_{-}x}
\left(
\begin{array}{l}
-e^{-i\varphi/2}\sin\frac{\theta}{2} \\
e^{i\varphi/2}\cos\frac{\theta}{2} 
\end{array}
\right)\;,
\label{wave_fn_region_2}
\end{eqnarray}
where $q_{\pm}=\sqrt{2m\left(V_{0}\pm\Gamma S-\epsilon\right)}/\hbar>0$. We choose $\Gamma<0$ such that ${\bf S}$ tends to align with ${\boldsymbol\sigma}$, and $q_{-}>q_{+}$. Since the effective mass $m$ merely rescales various energy parameters (see caption of Fig.~\ref{fig:fitting_Gr_gi_alphap}), it is set to be the same in the formalism below for simplicity.

{\it Spin-transfer torque.-} The amplitudes $A\sim E$ are solved by matching the wave function and its derivative at $x=0$, leaving only one free variable $|A|^{2}$ that is attributed to the Fermi surface-averaged spin density at the interface $|A|^{2}=N_{F}|{\boldsymbol\mu}_{0}|/a^{3}$, where $N_{F}$ is the density of states per $a^{3}$ at the Fermi surface, and $a$ is the lattice constant. The spin of conduction electrons inside the FMI can be calculated from the wave function in Eq.~(\ref{wave_fn_region_2}). It is customary\cite{Berger96} to express them in the frame $(x_{2},y_{2},z_{2})$ defined in Fig.~\ref{fig:fitting_Gr_gi_alphap} (a), where ${\hat{\bf z}_{2}}\parallel{\bf S}$, ${\hat{\bf y}_{2}}={\hat{\boldsymbol\mu}_{0}}\times{\hat{\bf S}}/\sin\theta$, and ${\hat{\bf x}_{2}}=-{\hat{\bf S}}\times\left({\hat{\boldsymbol\mu}_{0}}\times{\hat{\bf S}}\right)/\sin\theta$. The spinors in Eq.~(\ref{wave_fn_region_2}) are simply $(1\;0)^{T}$ and $(0\; 1)^{T}$ in this frame. The STT can be calculated by the following procedure\cite{Berger96}. Introducing  
\begin{eqnarray}
n_{\sigma\pm}&=&k_{0\sigma}/(k_{0\sigma}+iq_{\pm})\;,
\nonumber \\
\gamma_{\theta}&=&\frac{n_{\downarrow+}}{n_{\uparrow+}}\cos^{2}\frac{\theta}{2}
+\frac{n_{\downarrow-}}{n_{\uparrow-}}\sin^{2}\frac{\theta}{2}\approx 1\;,
\end{eqnarray} 
the conduction electron spin tunneled into the FMI is $\langle{\boldsymbol\sigma}\rangle=\langle\psi_{FI}|{\boldsymbol\sigma}|\psi_{FI}\rangle$, whose components are
\begin{eqnarray}
&&\langle\sigma^{x_{2},y_{2}}\rangle=-4\frac{|A|^{2}}{|\gamma_{\theta}|^{2}}\sin\theta e^{-\left(q_{+}+q_{-}\right)x}\left({\rm Re},{\rm Im}\right)\left(n_{\downarrow+}^{\ast}n_{\downarrow-}\right)\;,
\nonumber \\
&&\langle \sigma^{z_{2}}\rangle=2\frac{|A|^{2}}{|\gamma_{\theta}|^{2}}\cos\theta e^{-\left(q_{+}+q_{-}\right)x}
\nonumber \\
&&\;\;\;\times\left[\left(|n_{\downarrow+}|^{2}-|n_{\downarrow-}|^{2}\right)
+\left(|n_{\downarrow+}|^{2}+|n_{\downarrow-}|^{2}\right)\cos\theta\right]\;.
\label{SxSy_in_CMI}
\end{eqnarray}
The total spin per cross section channel is denoted by $\langle\overline{{\boldsymbol\sigma}}\rangle=a^{2}\int_{0}^{\infty}dx\langle{\boldsymbol\sigma}\rangle$.
The magnetization within the range of $\langle\overline{\boldsymbol\sigma}\rangle$, about $2\pi/\left(q_{+}+q_{-}\right)\sim a$, is treated as a macrospin ${\bf S}$. From the LL dynamics, the $s-d$ coupling $H_{sd}=\Gamma{\boldsymbol\sigma}\cdot{\bf S}$ renders the STT\cite{Berger96}, whose response in the damping-like and field-like direction define the spin mixing conductance\cite{Maekawa12} $G_{r}$ and $G_{i}$, respectively,
\begin{eqnarray}
{\boldsymbol\tau}&=&\frac{\Gamma}{\hbar}\langle\overline{{\boldsymbol\sigma}}\rangle\times{\bf S}=\frac{\Gamma}{\hbar}S\langle\overline{\sigma}^{y_{2}}\rangle{\hat{\bf x}_{2}}-\frac{\Gamma}{\hbar}S\langle\overline{\sigma}^{x_{2}}\rangle{\hat{\bf y}_{2}}
\nonumber \\
&=&\frac{\Gamma Sa^{2}N_{F}}{\hbar}\left[G_{r}{\hat{\bf S}}\times\left({\hat{\bf S}}\times{\boldsymbol\mu}_{0}\right)+G_{i}{\hat{\bf S}}\times{\boldsymbol\mu}_{0}\right]
\nonumber \\
G_{r,i}&=&\int_{0}^{\infty}\frac{\langle\sigma^{y_{2},x_{2}}\rangle}{N_{F}|{\boldsymbol\mu}_{0}|\sin\theta}dx
=-4\frac{\left({\rm Im},{\rm Re}\right)\left(n_{\downarrow+}^{\ast}n_{\downarrow-}\right)}{a^{3}|\gamma_{\theta}|^{2}\left(q_{+}+q_{-}\right)}\;,
\nonumber \\
\label{effective_EOM}
\end{eqnarray}
after substituting Eq.~(\ref{SxSy_in_CMI}) and $|A|^{2}=N_{F}|{\boldsymbol\mu}_{0}|/a^{3}$.

Alternatively, one can extract STT from the spin current\cite{Slonczewski96}. The spin current in the NM at position $x$ is
\begin{eqnarray}
{\boldsymbol j}_{x}=\frac{\hbar}{4im}\left[\psi_{N}^{\ast}{\boldsymbol\sigma}\left(\partial_{x}\psi_{N}\right)
-\left(\partial_{x}\psi_{N}^{\ast}\right){\boldsymbol\sigma}\psi_{N}\right]\;.
\label{spin_current_general}
\end{eqnarray} 
Conservation of angular momentum requires that the spin current at the interface to be equal to the STT, $a^{2}\left({\boldsymbol j}_{0}-{\boldsymbol j}_{\infty}\right)=a^{2}{\boldsymbol j}_{0}={\boldsymbol\tau}$\cite{Slonczewski96,Maekawa12,Chen13}, which is indeed satisfied in this approach. Consequently, one can use Eq.~(\ref{spin_current_general}) to extract $G_{r,i}$ \cite{supplementary_material}
\begin{eqnarray}
\frac{\Gamma SN_{F}}{\hbar}G_{r}&=&\frac{2j_{0}^{x}\cos\varphi}{|{\boldsymbol \mu}_{0}|\sin 2\theta}+\frac{2j_{0}^{y}\sin\varphi}{|{\boldsymbol \mu}_{0}|\sin 2\theta}=-\frac{j_{0}^{z}}{|{\boldsymbol\mu}_{0}|\sin^{2}\theta}\;,
\nonumber \\
\frac{\Gamma SN_{F}}{\hbar}G_{i}&=&\frac{j_{0}^{x}\sin\varphi}{|{\boldsymbol \mu}_{0}|\sin \theta}-\frac{j_{0}^{y}\cos\varphi}{|{\boldsymbol \mu}_{0}|\sin \theta}\;,
\label{mixing_from_Js}
\end{eqnarray}
which gives the same $G_{r,i}$ as in Eqs.~(\ref{effective_EOM}). In short, either extracting from the LL dynamics\cite{Berger96} in Eqs.~(\ref{effective_EOM}), or requiring angular momentum conservation\cite{Slonczewski96} and using Eq.~(\ref{spin_current_general}), one obtains the same STT. 


$G_{r}$ and $G_{i}$ are independent from the azimuthal angle $\varphi$, but have small $\theta$ dependence through $\gamma_{\theta}$ which can be considered as higher order contributions (At most few percent. Their ratio does not depend on $\theta$). The experimentally measurable quantity $G_{r,i}\times e^{2}/\hbar$ at $\theta=0.3\pi$ is plotted in Fig.~\ref{fig:fitting_Gr_gi_alphap} as a function of the $s-d$ coupling and insulating gap relative to the Fermi energy. Their ratio 
\begin{eqnarray}
\frac{G_{r}}{G_{i}}
=\frac{\int_{0}^{\infty}\langle\sigma^{y_{2}}\rangle dx}{\int_{0}^{\infty}\langle\sigma^{x_{2}}\rangle dx}=\frac{k_{0\downarrow}\left(q_{+}-q_{-}\right)}{k_{0\downarrow}^{2}+q_{+}q_{-}}\;.
\label{Onsager_relation}
\end{eqnarray}
strongly depends on the wave function tunneled into the FMI and is therefore highly influenced by the insulating gap and the $s-d$ coupling. In the supplementary material\cite{supplementary_material}, it is shown that the (anisotropic) spin relaxation, simulated phenomenologically by an exponentially decay factor\cite{Berger96}, is not crucial to determine the damping-like to field-like ratio in real materials. This is very different from the metallic heterostructures in which spin relaxation is the origin of STT and gives predominately a damping-like torque\cite{Berger96,Slonczewski96}. 



For a reasonable value of $s-d$ coupling, a very small ${\boldsymbol\mu}_{0}$ is already sufficient to create a damping-like torque that can overcome Gilbert damping and cause ferromagnetic resonance (FMR)\cite{Kajiwara10}, so the observed precession frequency (or the gyromagnetic ratio) is almost unaffected by the small field-like torque which can nevertheless be nonzero. A previous study shows that at the Ag/YIG interface, the enhanced local magnetic exchange field at the interface enhances $G_{r}$\cite{Jia11}, which can be interpreted as enhanced $s-d$ coupling in our calculation and is fairly consistent with the $\left\{q_{-},k_{0\downarrow}\right\}> q_{+}$ situation. In the Stoner limit $\left\{\Gamma,|{\boldsymbol \mu}_{0}|\right\}\ll\left\{\epsilon,V_{0}-\epsilon\right\}$, the field-like component dominates, in good agreement with Ref.\onlinecite{Jia11}.

{\it Spin pumping.-} The spin pumping at the NM/FMI interface can be addressed by solving the Bloch equation of conduction electrons in the presence of a dynamical magnetization\cite{Zhang04,Kajiwara10}
\begin{eqnarray}
\frac{\partial\langle{\boldsymbol\sigma}\rangle}{\partial t}+\partial_{x}{\boldsymbol j}_{x}=\frac{\Gamma}{\hbar}{\bf S}\times\langle{\boldsymbol\sigma}\rangle-\overline{\boldsymbol\Gamma}_{sf}
\label{Bloch_eq}
\end{eqnarray}
where $\overline{\boldsymbol\Gamma}_{sf}$ is a spin relaxation term. Assuming adiabatic process, i.e., the magnetization dynamics $|d{\bf S}/dt|$ is much slower than any frequency scale $\left\{\epsilon/\hbar,V_{0}/\hbar,\Gamma/\hbar\right\}$ in the problem, the wave functions in Eqs.~(\ref{NM_wave_function}) and (\ref{wave_fn_region_2}), as well as $\langle{\boldsymbol\sigma}\rangle$ in Eq.~(\ref{SxSy_in_CMI}), remain valid. As detailed in the supplementary material\cite{supplementary_material}, by considering the modification of $\langle{\boldsymbol\sigma}\rangle$ after a small time lapse $\delta t$ due to the magnetization dynamics and taking $\delta t\rightarrow 0$, the right hand side of Eq.~(\ref{Bloch_eq}) vanishes. Upon integrating over $x$ and using ${\boldsymbol j}_{\infty}=0$, Eq.~(\ref{Bloch_eq}) gives
\begin{eqnarray}
{\boldsymbol j}_{0}=\int_{0}^{\infty}\frac{\partial\langle{\boldsymbol\sigma}\rangle}{\partial t}dx=N_{F}|{\boldsymbol\mu}_{0}|\left[G_{r}{\hat{\bf S}}\times\frac{d{\hat{\bf S}}}{dt}+G_{i}\frac{d{\hat{\bf S}}}{dt}\right]\;,
\label{spin_pumping_eq}
\end{eqnarray}
where $G_{r,i}$ are the same as those obtained from Eqs.~(\ref{effective_EOM}) and (\ref{mixing_from_Js}). Here ${\boldsymbol\mu}_{0}$ is interpreted as a proximity induced spin accumulation. Comparing Eqs.~(\ref{effective_EOM}) and (\ref{spin_pumping_eq}), evidently the Onsager relation\cite{Maekawa12}, which dictates that the damping-like and field-like response for STT and for spin pumping must be equal, is satisfied.

A previous calculation\cite{Kajiwara10} gives generically $G_{i}=0$ in spin pumping, which is however at odds with the first principle calculation of STT that shows both $G_{r}$ and $G_{i}$ are nonzero\cite{Jia11}, as Onsager relation is not satisfied. Our calculation indicates that the material-dependent wave function is the key to resolve this discrepancy, and the spin pumping in general has both field-like and damping-like components. If the magnetization dynamics is a precession around a common axis, as that induced by FMR\cite{Kajiwara10}, the field-like component time-averages to zero while the damping-like component does not (see Fig.~1(b) of Ref.~\onlinecite{supplementary_material}), so the observed dc spin current may be well explained by the damping-like component alone which however does not rule out the possibility of nonzero field-like component. Moreover, the field-like component also contributes to the ac measurements\cite{Hahn13,Wei14,Weiler14} which identify a much larger ac spin current than the dc one. Thus none of the known experiments contradicts the possibility of having a field-like component.


\begin{figure}[ht]
\begin{center}
\includegraphics[clip=true,width=0.95\columnwidth]{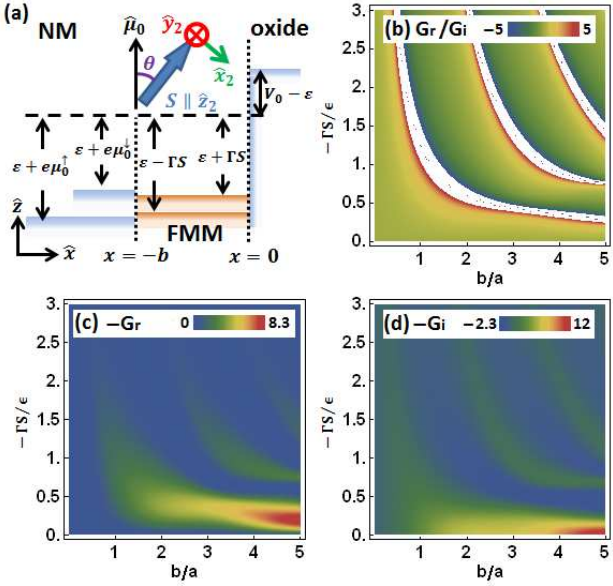}
\caption{(color online) Same as Fig.~(\ref{fig:fitting_Gr_gi_alphap}), but for NM/FMM/oxide trilayer. (b) to (d) are plotted in terms of the $s-d$ coupling relative to the Fermi energy $-\Gamma S/\epsilon$ and the thickness of the FMM layer $b/a$.  } 
\label{fig:NMFMMoxide_fitting_Gr_gi_alphap}
\end{center}
\end{figure}




{\it NM/FMM/oxide trilayer.-} The above analysis can also address the quantum effects in the NM/FMM/oxide trilayer, which may not be captured by diffusive approaches\cite{Vedyayev11,Manchon12}. Consider the trilayer shown in Fig.~\ref{fig:NMFMMoxide_fitting_Gr_gi_alphap}(a) that contains three regions: (1) A NM occupying $-\infty<x<-b$ described by Eq.~(\ref{NM_wave_function}) with the replacement $\mu_{0}^{\sigma}\rightarrow\mu_{-b}^{\sigma}$. (2) A FMM in $-b<x<0$ described by $H_{FM}=p^{2}/2m+\Gamma{\bf S}\cdot{\boldsymbol \sigma}$. It is unnecessary to include a Rashba term $\left({\bf p}\times{\bf E}_{SOC}\right)\cdot{\boldsymbol\sigma}$ because ${\bf p}\parallel{\bf E}_{SOC}\parallel{\hat{\bf x}}$ in this problem, where ${\bf E}_{SOC}$ is the effective electric field due to the parity-breaking in ${\hat{\bf x}}$-direction\cite{Manchon08,Pesin12,Haney10,Haney13}. We consider a FMM much thinner than its spin relaxation length $b\ll l_{sf}$ such that the wave function
\begin{eqnarray}
\psi_{FM}&=&\left(De^{ik_{+}x}+Fe^{-ik_{+}x}\right)
\left(
\begin{array}{l}
e^{-i\varphi/2}\cos\frac{\theta}{2} \\
e^{i\varphi/2}\sin\frac{\theta}{2} 
\end{array}
\right)
\nonumber \\
&+&\left(Ee^{ik_{-}x}+Ge^{-ik_{-}x}\right)
\left(
\begin{array}{l}
-e^{-i\varphi/2}\sin\frac{\theta}{2} \\
e^{i\varphi/2}\cos\frac{\theta}{2} 
\end{array}
\right)\;,
\label{wave_fn_FMM}
\end{eqnarray}
remains valid, where $k_{\pm}=\sqrt{2m(\epsilon\mp\Gamma S)}/\hbar$. (3) The oxide region at $x>0$ described by $H_{O}=p^{2}/2m+V_{0}\Theta(x)$, 
where $V_{0}>\epsilon$ is the potential step. The wave function is 
\begin{eqnarray}
\psi_{O}=He^{-qx}\left(
\begin{array}{l}
1 \\
0
\end{array}
\right)
+Ie^{-qx}\left(
\begin{array}{l}
0 \\
1
\end{array}
\right),
\label{oxide_wave_function}
\end{eqnarray}
where $q=\sqrt{2m(V_{0}-\epsilon)}/\hbar$. The wave functions render zero spin current at the FMM/oxide interface ${\boldsymbol j}_{0}=0$, but finite at the NM/FMM interface ${\boldsymbol j}_{-b}\neq 0$. After matching the wave functions, we introduce 
\begin{eqnarray}
n_{\alpha\beta}&=&\frac{k_{\alpha}}{k_{\alpha}+\beta iq}\;,
\;W_{\sigma\alpha\beta}=\frac{k_{0\sigma}+\beta k_{\alpha}}{2k_{0\sigma}}\;,
\nonumber \\
Z_{\sigma\alpha\beta}&=&W_{\sigma\alpha\beta}e^{-ik_{\alpha}b}+W_{\sigma\alpha\overline{\beta}}\frac{n_{\alpha+}}{n_{\alpha-}}e^{ik_{\alpha}b}\;,
\nonumber \\
\gamma_{\theta}^{\prime}&=&Z_{\uparrow++}Z_{\downarrow-+}\cos^{2}\frac{\theta}{2}+Z_{\downarrow++}Z_{\uparrow-+}\sin^{2}\frac{\theta}{2}\;,\;\;\;
\label{defined_variables}
\end{eqnarray}
the spin expectation values in the FMM are then
\begin{eqnarray}
&&\langle\sigma^{x_{2},y_{2}}\rangle=-\frac{|A|^{2}}{|\gamma_{\theta}^{\prime}|^{2}}\sin\theta\left\{{\rm Re},{\rm Im}\right\}\left\{Z_{\downarrow-+}^{\ast}Z_{\downarrow++}\right.
\nonumber \\
&&\times\left.\left[e^{-ik_{+}x}+\frac{n_{++}^{\ast}}{n_{+-}^{\ast}}e^{ik_{+}x}\right]
\left[e^{ik_{-}x}+\frac{n_{-+}}{n_{--}}e^{-ik_{-}x}\right]\right\}\;.\;\;\;\;\;\;\;
\label{spin_expectation_in_FMM}
\end{eqnarray}
The STT and spin pumping can be calculated from Eq.~(\ref{effective_EOM}) with the replacement $\int_{0}^{\infty}\langle\sigma^{y_{2},x_{2}}\rangle dx\rightarrow \int_{-b}^{0}\langle\sigma^{y_{2},x_{2}}\rangle dx$ and using Eq.~(\ref{spin_expectation_in_FMM}). The resulting $G_{r,i}$ has small $\theta$ dependence through $\gamma_{\theta}^{\prime}$ that can be attributed to higher harmonic terms allowed by symmetry\cite{Garello13}. The Onsager relation is again satisfied. Alternatively, STT can be calculated by requiring angular momentum conservation\cite{Slonczewski96} $a^{2}\left({\boldsymbol j}_{-b}-{\boldsymbol j}_{\infty}\right)=a^{2}{\boldsymbol j}_{-b}={\boldsymbol\tau}$ and using Eqs.~(\ref{spin_current_general}) and (\ref{mixing_from_Js}) with the replacement ${\boldsymbol j}_{0}\rightarrow{\boldsymbol j}_{-b}$, which yields the same $G_{r,i}$.

The numerical results for $G_{r,i}$ at $\theta=0.3\pi$ are shown in Fig.~\ref{fig:NMFMMoxide_fitting_Gr_gi_alphap}, where certain oscillations with respect to the $s-d$ coupling and the FMM layer thickness are evident, signaturing the effect of quantum interference, while $G_{r,i}$ do not strongly depend on the insulating gap. This is in accordance with the measurement in Ta/CoFeB/MgO that shows hints for a varying $G_{r}/G_{i}$ when changing FMM thickness\cite{Kim13}, although one should keep in mind that we consider the NM thickness to be much larger than its spin relaxation length so the injected spin current saturates\cite{Liu12,Liu11}, and hence our result does not depend on NM thickness unlike the experimental case. The absolute magnitude of $|G_{r,i}|\sim 10^{-15}\sim 10^{-14}\Omega^{-1}m^{-1}$ is close to that observed experimentally, thus the SHE-STT contribution relative to the SOT should not be overlooked.

In summary, the generic material properties in heterostructures involving insulators, such as $s-d$ coupling, insulating gap, and thickness of the ferromagnet, are shown to strongly influence the tunneling wave function and subsequently the damping-like and field-like component of the STT and the spin pumping caused by SHE. Spin relaxation, on the other hand, plays a relatively minor role. The quantum effects of the wave function, such as quantum tunneling and quantum interference, are quantified in our minimal model that incorporates simultaneously the LL dynamics, angular momentum conservation, and Onsager relation, thus a convincing and inexpensive model to guide the search for appropriate materials. 


We thank P. W. Brouwer, P. Gambardella, P. Horsch, S. Maekawa, M. Mori, T. S. Nunner, J. Mendil, Y. Tserkovnyak, G. Vignale, and H.-H. Lin for stimulating discussions.

\begin{eqnarray}
\hline\nonumber
\end{eqnarray}
{\bf Supplementary Material}
\begin{eqnarray}
\hline\nonumber
\end{eqnarray}

\section{Detail of spin-transfer torque calculation }

\subsection{NM/FMI bilayer}

By matching the wave function in Eq.~(1) and (2) of the main text and their derivative at the interface $x=0$, the scattering coefficients in the NM/FMI bilayer read
\begin{eqnarray}
B&=&A\left[\frac{2n_{\downarrow+}\cos^{2}\frac{\theta}{2}}{\gamma_{\theta}}
+\frac{2n_{\downarrow-}\sin^{2}\frac{\theta}{2}}{\gamma_{\theta}}-1\right]\;,
\nonumber \\
C&=&Ae^{i\varphi}\frac{\left(n_{\downarrow+}-n_{\downarrow-}\right)}{\gamma_{\theta}}\sin\theta\;,
\nonumber \\
D&=&2Ae^{i\varphi/2}\frac{n_{\downarrow+}}{\gamma_{\theta}}\cos\frac{\theta}{2}\;,
\nonumber \\
E&=&-2Ae^{i\varphi/2}\frac{n_{\downarrow-}}{\gamma_{\theta}}\sin\frac{\theta}{2}\;,
\end{eqnarray}
where $n_{\sigma\pm}$ and $\Gamma_{\theta}$ are defined in Eq.~(3) of the main text. Since the particle flux is zero $k_{0\uparrow}|A|^{2}-k_{0\uparrow}|B|^{2}-k_{0\downarrow}|C|^{2}=0$, there is no charge current in this problem but only spin current, making it clear that the formalism describes the spin injection caused by the pure spin current in SHE.

To extract the spin mixing conductance $G_{r,i}$ from angular momentum conservation, we identify spin current at $x=0$ with the STT
\begin{eqnarray}
{\boldsymbol j}_{0}&=&\left.\frac{\hbar}{4im}\left[\psi_{N}^{\ast}{\boldsymbol\sigma}\left(\partial_{x}\psi_{N}\right)
-\left(\partial_{x}\psi_{N}^{\ast}\right){\boldsymbol\sigma}\psi_{N}\right]\right|_{x=0}
\nonumber \\
&=&\frac{\boldsymbol\tau}{a^{2}}=\frac{\Gamma SN_{F}}{\hbar}\left[G_{r}{\hat{\bf S}}\times\left({\hat{\bf S}}\times{\boldsymbol\mu}_{0}\right)+G_{i}{\hat{\bf S}}\times{\boldsymbol\mu}_{0}\right]
\nonumber \\
&=&\frac{\Gamma SN_{F}|{\boldsymbol\mu}_{0}|}{\hbar}\left\{\sin\theta\left[G_{r}\cos\theta\cos\varphi+G_{i}\sin\varphi\right]{\hat{\bf x}}\right.
\nonumber \\
&&\left.+\sin\theta\left[G_{r}\cos\theta\sin\varphi-G_{i}\cos\varphi\right]{\hat{\bf y}}-G_{r}\sin^{2}\theta\;{\hat{\bf z}}\right\}\;, \nonumber \\
\end{eqnarray} 
where we have used ${\hat{\bf S}}\times\left({\hat{\bf S}}\times{\boldsymbol\mu}_{0}\right)=|{\boldsymbol\mu}_{0}|\;{\hat{\bf S}}\times\left({\hat{\bf S}}\times{\hat{\bf z}}\right)$ and ${\hat{\bf S}}\times{\boldsymbol\mu}_{0}=|{\boldsymbol\mu}_{0}|\;{\hat{\bf S}}\times{\hat{\bf z}}$ since the spin voltage is defined to be polarized along ${\boldsymbol\mu}_{0}\parallel {\hat{\bf z}}=(0,0,1)$, and the fact that the magnetization points at ${\hat{\bf S}}=(\cos\theta\cos\varphi,\cos\theta\sin\varphi,\sin\theta)$. Solving for $G_{r,i}$ in terms of each component of the spin current yields Eq.~(7) of the main text.

We now address the issue of spin relaxation, which in general can be anisotropic. Since we are not aware of any convincing calculation of spin relaxation for the evanescent wave function tunneled into the FMI, we model the spin relaxation phenomenologically by an exponentially decayed factor on the spin expectation value, although our argument below does not require an exponential form. From Eq.~(4) in the main text, adding anisotropic spin relaxation length $\left\{\lambda_{x_{2}},\lambda_{y_{2}}\right\}$ into the spin expectation value yields
\begin{eqnarray}
\langle\sigma^{x_{2},y_{2}}\rangle&=&-4\frac{|A|^{2}}{|\gamma_{\theta}|^{2}}\sin\theta e^{-x/\lambda_{x_{2},y_{2}}}e^{-\left(q_{+}+q_{-}\right)x} 
\nonumber \\
&&\times\left({\rm Re},{\rm Im}\right)\left(n_{\downarrow+}^{\ast}n_{\downarrow-}\right)\;,
\label{SxSy_in_CMI}
\end{eqnarray}
which gives Eq.~(5) in the main text with corrections
\begin{eqnarray}
G_{r,i}&=&\int_{0}^{\infty}\frac{\langle\sigma^{y_{2},x_{2}}\rangle}{N_{F}|{\boldsymbol\mu}_{0}|\sin\theta}dx
\nonumber \\
&=&-4\frac{\left({\rm Im},{\rm Re}\right)\left(n_{\downarrow+}^{\ast}n_{\downarrow-}\right)}{a^{3}|\gamma_{\theta}|^{2}\left[\left(q_{+}+q_{-}\right)+1/\lambda_{y_{2},x_{2}}\right]}\;,
\nonumber \\
&=&-4\frac{\left({\rm Im},{\rm Re}\right)\left(n_{\downarrow+}^{\ast}n_{\downarrow-}\right)}{a^{3}|\gamma_{\theta}|^{2}\left(1/\xi_{pen}+1/\lambda_{y_{2},x_{2}}\right)}\;,
\end{eqnarray}
rendering the $G_{r}$ and $G_{i}$ that depend on $\lambda_{y_{2}}$ and $\lambda_{x_{2}}$, respectively, so the damping-like to field-like ratio $G_{r}/G_{i}$ is influenced by the spin relaxation if it is anisotropic $\lambda_{x_{2}}\neq\lambda_{y_{2}}$. However, one sees that in the above equation, the more detrimental factor is the penetration depth of the evanescent wave function $\xi_{pen}=1/(q_{+}+q_{-})\sim \hbar/\sqrt{2m(V_{0}-\epsilon)}$ determined by the insulating gap $V_{0}-\epsilon$. The size of insulating gap in oxide insulators is typically of eV, yielding $\xi_{pen}$ of the order of lattice constant, so it is unlikely that $\left\{\lambda_{x_{2}},\lambda_{y_{2}}\right\}$ can be smaller than $\xi_{pen}$. In other words, unless the anisotropy of spin relaxation is associated with an energy scale (spin-orbit coupling, impurities, interface roughness, etc, which are beyond the scope of this approach) that is comparable to the insulating gap, the anisotropy is not crucial. With the assumption that $\xi_{pen}\ll\left\{\lambda_{x_{2}},\lambda_{y_{2}}\right\}$ and a clean interface, from Eq.~(4) one sees that the influence of $\{\lambda_{x_{2}},\lambda_{y_{2}}\}$ on $G_{r}/G_{i}$ is negligible. Physically, this means that the electron can only penetrate FMI over a short distance within which its spin has yet started to relax.

\subsection{NM/FMM/oxide trilayer}

By matching the wave functions in Eq.~(1), (11), and (12) in the main text and their derivative in the two interfaces $x=-b$ and $x=0$, the scattering coefficients in NM/FMM/oxide trilayer are
\begin{eqnarray}
B&=&A\;\frac{e^{-2ik_{0\uparrow}b}}{\gamma_{\theta}^{\prime}}
\left[Z_{\uparrow+-}Z_{\downarrow-+}\cos^{2}\frac{\theta}{2}+Z_{\uparrow--}Z_{\downarrow++}\sin^{2}\frac{\theta}{2}\right]\;,
\nonumber \\
C&=&A\;\frac{e^{-i\left(k_{0\uparrow}+k_{0\downarrow}\right)b+i\varphi}}{2\gamma_{\theta}^{\prime}}
\left[Z_{\downarrow+-}Z_{\downarrow-+}-Z_{\downarrow--}Z_{\downarrow++}\right]\sin\theta\;,
\nonumber \\
D&=&A\;\frac{e^{-ik_{0\uparrow}b+i\varphi/2}}{\gamma_{\theta}^{\prime}}Z_{\downarrow-+}\cos\frac{\theta}{2}\;,\;F=\frac{n_{++}}{n_{+-}}D\;,
\nonumber \\
E&=&-A\;\frac{e^{-ik_{0\uparrow}b+i\varphi/2}}{\gamma_{\theta}^{\prime}}Z_{\downarrow++}\sin\frac{\theta}{2}\;,\;G=\frac{n_{-+}}{n_{--}}E\;,
\nonumber \\
H&=&2A\;\frac{e^{-ik_{0\uparrow}b}}{\gamma_{\theta}^{\prime}}
\left[n_{++}Z_{\downarrow-+}\cos^{2}\frac{\theta}{2}+n_{-+}Z_{\downarrow++}\sin^{2}\frac{\theta}{2}\right]\;,
\nonumber \\
I&=&A\;\frac{e^{-ik_{0\uparrow}b+i\varphi}}{\gamma_{\theta}^{\prime}}
\left[n_{++}Z_{\downarrow-+}-n_{-+}Z_{\downarrow++}\right]\sin\theta\;,
\end{eqnarray}
where $n_{\alpha\beta}$, $Z_{\sigma\alpha\beta}$, and $\gamma_{\theta}^{\prime}$ are those defined in Eq.~(13) of the main text.

\section{Detail of spin pumping calculation}

\begin{figure}[ht]
\begin{center}
\includegraphics[clip=true,width=0.95\columnwidth]{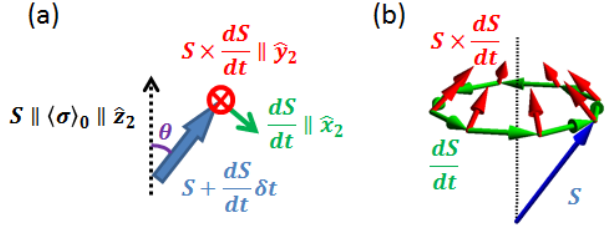}
\caption{(color online) (a) Schematics of the magnetization dynamics and the coordinate after time lapse $\delta t$. (b) The magnetization dynamics induced by FMR, where the field-like component (green arrows) time-averages to zero, while the damping-like component (red arrows) does not. } 
\label{fig:spin_pumping_at_delta_t}
\end{center}
\end{figure}

The spin pumping can be calculated from our tunneling formalism using adiabatic assumption, i.e., the magnetization dynamics is much slower than any characteristic frequency scales in the problem. We start from the Bloch equation described by Eq.~(9) in the main text and use NM/FMI bilayer as an example. When the magnetization ${\bf S}$ is static, the conduction electrons leaked from NM into FMI are spin-split because of the exchange field $\Gamma{\bf S}\cdot{\boldsymbol\sigma}$, causing an equilibrium spin density that can be described by Eq.~(4) of the main text, where the incoming flux $|A|^{2}=N_{F}|{\boldsymbol\mu}_{0}|/a^{3}$ and the spin voltage at the interface ${\boldsymbol\mu}_{0}$ are regarded as phenomenological parameters to fit the equilibrium spin density. The magnetization dynamics modifies this equilibrium spin density as we now address. Suppose at $t=0$, the conduction electron spin is in equilibrium with the magnetization, so $\theta=0$ and 
\begin{eqnarray}
\langle{\boldsymbol\sigma}(x,0)\rangle=\langle{\boldsymbol\sigma}(x,0)\rangle_{0}\parallel{\bf S}\parallel{\hat{\bf z}_{2}}\;.
\end{eqnarray}
Because of the magnetization dynamics  $d{\bf S}/dt$, within an infinitely small time lapse $\delta t$, the conduction electron spin is modified by
\begin{eqnarray}
\langle{\boldsymbol\sigma}(x,\delta t)\rangle=\langle{\boldsymbol\sigma}(x,0)\rangle_{0}+\delta\langle{\boldsymbol\sigma}(x,\delta t)\rangle\;.
\end{eqnarray}
If $d{\bf S}/dt$ is much slower than any characteristic frequency scale $\left\{\epsilon/\hbar,V_{0}/\hbar,\Gamma/\hbar\right\}$ in the problem, then one can assume that the system remains quasi-equilibrium so the wave functions we used are still applicable. This is generally true in present experiments, since the magnetization dynamics, usually generated by ferromagnetic resonance (FMR), is typically $\sim$GHz, while the fermionic frequency scales $\left\{\epsilon/\hbar,V_{0}/\hbar,\Gamma/\hbar\right\}$ are usually above THz. Since $d{\bf S}/dt$ is pulling ${\bf S}$ away from $\langle{\boldsymbol\sigma}(x,0)\rangle_{0}$, it is along ${\hat{\bf x}_{2}}$, and ${\bf S}\times d{\bf S}/dt$ is along ${\hat{\bf y}_{2}}$, as shown in Fig.~\ref{fig:spin_pumping_at_delta_t} (a). In other words, 
\begin{eqnarray}
{\hat{\bf x}_{2}}&=&\frac{1}{\left|\frac{d{\bf S}}{dt}\right|}\left(\frac{d{\bf S}}{dt}\right)\;,
\nonumber \\
{\hat{\bf y}_{2}}&=&\frac{1}{\left|\frac{d{\bf S}}{dt}\right|}\left({\hat{\bf S}}\times\frac{d{\bf S}}{dt}\right)\;,
\label{pumping_coordinate}
\end{eqnarray}
The infinitely small angle developed between ${\bf S}$ and $\langle{\boldsymbol\sigma}(x,0)\rangle_{0}$ after time lapse $\delta t$ is 
\begin{eqnarray}
\theta=\left|\frac{d{\hat{\bf S}}}{dt}\right|\delta t=\frac{1}{S}\left|\frac{d{\bf S}}{dt}\right|\delta t\;,
\end{eqnarray}
therefore
\begin{eqnarray}
&&\lim_{\delta t\rightarrow 0}\sin\theta=\left|\frac{d{\hat{\bf S}}}{dt}\right|\delta t=\frac{1}{S}\left|\frac{d{\bf S}}{dt}\right|\delta t\;,
\nonumber \\
&&\lim_{\delta t\rightarrow 0}\cos\theta=1\;.
\label{angle_developed}
\end{eqnarray}
The modification of spin developed after the small time lapse $\delta t$ is, using Eq.~(4) in the main text (ignoring spin relaxation as argued in Sec.~1.1) together with Eqs.~(\ref{pumping_coordinate}) and (\ref{angle_developed}),
\begin{eqnarray}
\delta\langle{\boldsymbol\sigma}\rangle&=&\langle\sigma^{x_{2}}\rangle{\hat{\bf x}_{2}}+\langle\sigma^{y_{2}}\rangle{\hat{\bf y}_{2}}
\nonumber \\
&=&-4\frac{|A|^{2}}{|\gamma_{\theta}|^{2}}\frac{1}{S}\left|\frac{d{\bf S}}{dt}\right|\delta te^{-\left(q_{+}+q_{-}\right)x}
\nonumber \\
&&\times\left[{\hat{\bf x}_{2}}{\rm Re}\left(n_{\downarrow+}^{\ast}n_{\downarrow-}\right)+{\hat{\bf y}_{2}}{\rm Im}\left(n_{\downarrow+}^{\ast}n_{\downarrow-}\right)\right]
\nonumber \\
&=&-4\frac{|A|^{2}}{|\gamma_{\theta}|^{2}}\delta te^{-\left(q_{+}+q_{-}\right)x}
\nonumber \\
&&\times\left[\frac{d{\hat{\bf S}}}{dt}{\rm Re}\left(n_{\downarrow+}^{\ast}n_{\downarrow-}\right)+{\hat{\bf S}}\times\frac{d{\hat{\bf S}}}{dt}{\rm Im}\left(n_{\downarrow+}^{\ast}n_{\downarrow-}\right)\right]\;.
\nonumber \\
\end{eqnarray}
So the time derivative in the Bloch equation is independent of $\delta t$
\begin{eqnarray}
\frac{\partial{\boldsymbol\sigma}}{\partial t}&=&\frac{\delta{\boldsymbol\sigma}}{\delta t}
\nonumber \\
&=&-4\frac{|A|^{2}}{|\gamma_{\theta}|^{2}}e^{-\left(q_{+}+q_{-}\right)x}
\nonumber \\
&&\times\left[\frac{d{\hat{\bf S}}}{dt}{\rm Re}\left(n_{\downarrow+}^{\ast}n_{\downarrow-}\right)+{\hat{\bf S}}\times\frac{d{\hat{\bf S}}}{dt}{\rm Im}\left(n_{\downarrow+}^{\ast}n_{\downarrow-}\right)\right]
\nonumber \\
&=&\frac{\langle\sigma^{x_{2}}\rangle}{\sin\theta}\frac{d{\hat{\bf S}}}{dt}+\frac{\langle\sigma^{y_{2}}\rangle}{\sin\theta}{\hat{\bf S}}\times\frac{d{\hat{\bf S}}}{dt}\;.
\end{eqnarray}
Similarly, the other two terms in the Bloch equation are 
\begin{eqnarray}
&&\frac{\Gamma}{\hbar}{\bf S}\times\langle{\boldsymbol\sigma}\rangle=\frac{\delta t}{\tau_{sd}}\left[-\frac{d{\hat{\bf S}}}{dt}\;\frac{\langle\sigma^{y_{2}}\rangle}{\sin\theta}+{\hat{\bf S}}\times\frac{d{\hat{\bf S}}}{dt}\;\frac{\langle\sigma^{x_{2}}\rangle}{\sin\theta}\right]\;,
\nonumber \\
&&\overline{\boldsymbol\Gamma}_{sf}=\frac{\delta\langle{\boldsymbol\sigma}\rangle}{\tau_{sf}}
=\frac{\delta t}{\tau_{sf}}\left[\frac{d{\hat{\bf S}}}{dt}\;\frac{\langle\sigma^{x_{2}}\rangle}{\sin\theta}+{\hat{\bf S}}\times\frac{d{\hat{\bf S}}}{dt}\;\frac{\langle\sigma^{y_{2}}\rangle}{\sin\theta}\right]\;,\;\;\;
\nonumber \\
\end{eqnarray}
which are proportional to $\delta t$ and hence vanish at $\delta t\rightarrow 0$. This also indicates that both the spin relaxation time $\tau_{sf}$ and the $s-d$ interaction time scale $\tau_{sd}=\hbar/\Gamma S$ are not detrimental to the spin pumping mechanism. The Bloch equation is then reduced to the continuity equation 
\begin{eqnarray}
\frac{\partial\langle{\boldsymbol\sigma}\rangle}{\partial t}+\partial_{x}{\boldsymbol j}_{x}=0\;.\;\;\;\;\left({\rm at}\;\delta t\rightarrow 0\right)
\end{eqnarray}
By integrating over $x$ and using ${\boldsymbol j}_{\infty}=0$, one obtains Eq.~(10) of the main text. The spin pumping in the NM/FMM/oxide trilayer follows the same argument with the proper adaption of wave function and spin accumulation therein, and one arrives at the same Eq.~(10).

\end{document}